\documentstyle[12pt]{article}
\catcode`@=11
\newcount\@tempcntc
\def\@citex[#1]#2{\if@filesw\immediate\write\@auxout{\string\citation{#2}}\fi
  \@tempcnta\z@\@tempcntb\m@ne\def\@citea{}\@cite{\@for\@citeb:=#2\do
    {\@ifundefined
       {b@\@citeb}{\@citeo\@tempcntb\m@ne\@citea\def\@citea{,}{\bf ?}\@warning
       {Citation `\@citeb' on page \thepage \space undefined}}%
    {\setbox\z@\hbox{\global\@tempcntc0\csname b@\@citeb\endcsname\relax}%
     \ifnum\@tempcntc=\z@ \@citeo\@tempcntb\m@ne
       \@citea\def\@citea{,}\hbox{\csname b@\@citeb\endcsname}%
     \else
      \advance\@tempcntb\@ne
      \ifnum\@tempcntb=\@tempcntc
      \else\advance\@tempcntb\m@ne\@citeo
      \@tempcnta\@tempcntc\@tempcntb\@tempcntc\fi\fi}}\@citeo}{#1}}
\def\@citeo{\ifnum\@tempcnta>\@tempcntb\else\@citea\def\@citea{,}%
  \ifnum\@tempcnta=\@tempcntb\the\@tempcnta\else
   {\advance\@tempcnta\@ne\ifnum\@tempcnta=\@tempcntb \else \def\@citea{--}\fi
    \advance\@tempcnta\m@ne\the\@tempcnta\@citea\the\@tempcntb}\fi\fi}
\catcode`@=12

\def\barr{\begin{array}}
\def\earr{\end{array}}
\def\beq{\begin{equation}}
\def\eeq{\end{equation}}
\def\bea{\begin{eqnarray}}
\def\eea{\end{eqnarray}}
\def\la{\lambda_\alpha}
\def\bmath{\begin{displaymath}}
\def\emath{\end{displaymath}}
\def\bq{\begin{quote}}
\def\eq{\end{quote}}

\def\cL{{\cal L}}

\def\lN{\lambda_N}
\def\lt{\lambda_t}
\def\li{\lambda_i}

\def\PL{\mbox{P}_L}
\def\PR{\mbox{P}_R}
\def\apprle{\hspace{-0.1cm}\stackrel{\displaystyle <}{\sim}}
\def\apprge{\hspace{-0.1cm}\stackrel{\displaystyle >}{\sim}}
\def\slash#1{\setbox0=\hbox{$#1$}#1\hskip-\wd0\hbox to\wd0{\hss\sl/\/\hss}}

\def\mpla#1{{\em Mod.\ Phys.\ Lett.\ }{\bf A#1}}

\def\npb#1{{\em Nucl.\ Phys.\ }{\bf B#1}}
\def\plb#1{{\em Phys.\ Lett.\ }{\bf A#1}}
\def\plb#1{{\em Phys.\ Lett.\ }{\bf B#1}}
\def\prl#1{{\em Phys.\ Rev.\ Lett.\ }{\bf #1}}

\def\prd#1{{\em Phys.\ Rev.\ }{\bf D#1}}

\def\ptp#1{{\em Prog.\ Theor.\ Phys.\ }{\bf #1}}

\def\zpc#1{{\em Z.\ Phys.\ }{\bf C#1}}
\begin{document}

\begin{flushright}
RAL/94-090\\[-0.2cm]
MZ-TH/94-22\\[-0.2cm]
August 1994
\end{flushright}

\begin{center}
{\bf{\Large Heavy-Neutrino Chirality Enhancement of the Decay}}\\[0.3cm]
{\bf{\Large{\boldmath $K_L \to e\mu $} in Left-Right Symmetric
Models}}\\[2.5cm]
{\large Z.~Gagyi-Palffy}$^a$\footnote[1]{Supported by Graduiertenkolleg
Teilchenphysik, Mainz, Germany.\\ ~~~~~~~E-mail address:
gagyi\_palffy@vipmzw.physik.uni-mainz.de}{\large
,~A.~Pilaftsis}$^b$\footnote[2]{E-mail address: pilaftsis@v2.rl.ac.uk}{\large
,~and~K.~Schilcher}$^a$\\[0.4cm]
$^a${\it Institut f\"ur Physik, THEP, Johannes Gutenberg-Universit\"at,
55099 Mainz, Germany}\\[0.3cm]
$^b${\it Rutherford Appleton Laboratory, Chilton, Didcot, Oxon, OX11 0QX,
UK}
\end{center}
\vskip2cm
\centerline{\bf ABSTRACT}

We study the decay $K_L\to e\mu$ in minimal extensions
of the Standard Model based on the gauge groups $SU(2)_L
\otimes U(1)_Y$ and $SU(2)_R\otimes SU(2)_L\otimes U(1)_{B-L}$,
in which heavy Majorana neutrinos are present. In
$SU(2)_L\otimes U(1)_Y$ models with chiral neutral singlets,
$B(K_L\to e\mu )$ cannot be much larger than $5\times
10^{-15}$ without violating other low-energy constraints.
In $SU(2)_L\otimes SU(2)_R\otimes U(1)_{B-L}$ models, we find
that heavy-neutrino-chirality enhancements due to the
presence of left-handed and right-handed currents can give
rise to a branching ratio close to the present experimental
limit $B(K_L\to e \mu)<3.3\times 10^{-11}$.

\newpage

One of the salient features of the minimal Standard Model (SM)
is that the separate leptonic quantum numbers are conserved
to all orders of perturbation theory. However, if the SM is
considered to be the low-energy limit of a more fundamental
theory ({\em e.g.}, superstrings or grand unified theories), one may
then have to worry about large flavour-changing-neutral current
(FCNC) effects that could violate experimental data.
Among the possible FCNC decays forbidden in the SM,
the decay $K_L\to e\mu$ can play
a central r\^ole either to constrain or establish new physics beyond
the SM. Furthermore, it is already known that extensions of the SM
containing more than one neutral isosinglet can dramatically relax the
severe constraints on the mixings between light and heavy
neutrinos~\cite{WW,EW,BW,HN},
which are dictated by usual see-saw scenarios~\cite{seesaw}.
Such models with enhanced light-heavy neutrino mixings are also
associated with large Dirac mass terms in the general neutrino mass
matrix~\cite{HN}. As an immediate phenomenological consequence, nondecoupling
virtual effects originating from heavy neutrinos can considerably enhance the
decay rates  $H\to l\bar{l'}$~\cite{APetal}, $Z\to l\bar{l'}$~\cite{KPS},
$\tau \to eee$~\cite{IP}, and the values of other observables~\cite{BKPS}
to an experimentally accessible level.

In this note, we will investigate similar nondecoupling effects coming
from heavy neutrinos and the heavy top quark in the decay $K_L\to e\mu$.
We will analyze such effects in Majorana neutrino models based on the
gauge groups $SU(2)_L\otimes U(1)_Y$ and
$SU(2)_R\otimes SU(2)_L \otimes U(1)_{B-L}$~\cite{PS,MS,SV}.
In particular, in general $SU(2)_R\otimes SU(2)_L \otimes U(1)_{B-L}$ models
with no manifest or pseudomanifest left-right symmetry, a significant
enhancement occurs due to the  chirality difference between left-handed and
right-handed currents, leading to observable rates for the decay $K_L\to e\mu$.
The latter may be related to the observation made by Paschos
in~\cite{Paschos} for the $K_L-K_S$ mass difference in left-right
symmetric models.

In order to briefly describe the electroweak sector of the SM with one
right-handed neutrino per family, we will adopt the notations given
in~\cite{HN}.
Later on, we will extend our analysis to the
$SU(2)_L\otimes SU(2)_R\otimes U(1)_{B-L}$ gauge group.
To be specific, we will first consider an $SU(2)_L\otimes U(1)_Y$
symmetric model with $n_G$ generations of charged leptons $l_i$
($i=1,\ldots\; ,n_G$) and light (heavy) Majorana neutrinos $\nu_\alpha$
($N_\alpha$). The charged-current interaction of this models
is then governed by the Lagrangian
\beq
\cL^W_{int}\  =\  -\, \frac{g_w}{\sqrt{2}} W^{-\mu}
\sum_{i=1}^{n_G} \sum_{\alpha=1}^{2n_G} \overline{l}_i \gamma_{\mu} \PL
B_{l_i\alpha}n_\alpha\quad +\quad H.c., \label{WLint}
\eeq 
where $g_w$ is the $SU(2)_L$ weak coupling constant,
$\PL\ (\PR) =(1 - (+) \gamma_5)/2$, and $B$ is an $n_G \times 2n_G$
matrix obeying a number of useful identities given in~\cite{HN}.
In Eq.~(\ref{WLint}), we have collectively defined the mass eigenstates
of the light and heavy neutrinos as follows:
$n_\alpha\equiv\nu_\alpha$ for $\alpha=1,\ldots\; ,n_G$ and
$n_\alpha\equiv N_{\alpha -n_G}$ for
$\alpha =n_G+1,\ldots\; ,2n_G$.

In the model under consideration, the decay
$K_L \to e\mu$ is induced by the diagrams shown in Fig. 1(a)--(d).
In Fig.~1(b)--(d), the field $\chi_L$ describes the would-be Goldstone boson
in the Feynman--'t Hooft gauge, which is related to the longitudinal
polarization of the $W$ boson
in the unitary gauge.
In the applicable limit of vanishing external momenta, the amplitude of
the decay process $K_L\to e\mu$ takes the general form
\bea
A & = & \left( \frac{g_w}{\sqrt{2}}\right) ^4
\left< 0\left| \overline{d} \gamma_{\kappa} \PL s \right|
\overline{K} ^0\right> \overline{u}_{\mu} \gamma^{\kappa} \PL
v_e \frac{1}{(4\pi )^2} \frac{1}{M_{W}^2}\nonumber \\
& & \times \sum_{u_i=u,c,t} V_{id}^{*} V_{is}
  \sum_{\alpha = 1}^{2n_G} B_{\mu\alpha} B^{*}_{e\alpha} I(\li,\la),\label{AL}
\eea  
where $\la = m_{n_\alpha}^{2}/M^{2}_{W}$,
$\li = m_{u_i}^{2}/M^{2}_{W}$, $V$ is the usual
Cabbibo-Kobayashi-Maskawa (CKM) matrix, and the loop function
$I$ is obtained from the Feynman graphs in Fig.~1(a)--(d)~\cite{Inami}.
The function $I$ is analytically given by
\beq
I(\li,\la) = \left( 1+\frac 14\li\la \right) I_1(\li,\la)+
      2\li\la I_2(\li,\la),
\eeq 
with
\bea
I_1(\li,\la)  &=&
\left[ \frac{\li^2\ln \li}{(\la-\li)(1-\li)^2}
   +\frac{\la^2\ln \la}{(\li-\la)(1-\la)^2}
   -\frac{1}{(1-\li)(1-\la)}\right] ,\\
I_2(\li,\la)  &=&   -\ \left[
\frac{\li\ln \li}{(\la-\li)(1-\li)^2}
   +\frac{\la\ln \la}{(\li-\la)(1-\la)^2}
   +\frac{1}{(1-\li)(1-\la)}\right] .
\eea 
Following closely Ref.~\cite{PRD}, we define a reduced amplitude
$\tilde A$ through the expression
\beq
A = \left( \frac{g_w}{\sqrt{2}}\right) ^4
\left< 0\left| \overline{d} \gamma_{\kappa} \PL s \right|
\overline{K} ^0\right> \overline{u}_{\mu} \gamma^{\kappa} \PL
v_e \frac{1}{(4\pi )^2} \frac{1}{M_{W}^2}\tilde A,
\eeq 
where
\beq \tilde A = \sum_{i=u,c,t} V^*_{id} V_{is}
\sum_{\alpha=1}^{2n_G} B_{\mu\alpha} B^{*}_{e\alpha} I(\li,\la)\, .
\label{Ared}
\eeq 
The advantage of this definition is that $\tilde{A}$ is
a dimensionless quantity carrying the whole electroweak physics.
The matrix $B$ obeys a generalized GIM identity~\cite{GIM}
\beq
\sum_{\alpha=1}^{2n_G}B_{l_1\alpha}B_{l_2\alpha}^{*}=\delta_{l_1l_2},
\eeq 
similar to the known one satisfied by the CKM matrix $V$.
A double GIM mechanism is then operative both for the intermediate
$u$-type quarks and neutral leptons, which simplifies~Eq.~(\ref{Ared}) to
\beq
\tilde A = \sum_{i=c,t} V_{id}^{*} V_{is}
\sum_{\alpha=n_G+1}^{2n_G} B_{\mu \alpha} B^{*}_{e\alpha}
E(\li,\la),\label{red}
\eeq 
with
\bea
E(\li,\la)&=& \li\la \left\{
-\frac 34\frac{1}{(1-\li)(1-\la)}
\right.\nonumber\\
& &+ \left[ \frac 14- \frac 32 \frac{1}{\li-1}-\frac 34\frac{1}{(\li-1)^2}
\right] \frac{\ln\li}{\li-\la}\nonumber \\
& & +\left.
\left[ \frac 14- \frac 32 \frac{1}{\la-1}-\frac 34\frac{1}{(\la-1)^2}
\right] \frac{\ln\la}{\la-\li}\right\}. \label{Eps}
\eea 
In Eq.~(\ref{red}), we have considered that the up quark $u$ and the
light neutrinos $\nu_e$, $\nu_\mu$, and $\nu_\tau$
are massless. Only the virtual $c$ and $t$ quarks, and
the $n_G$ heavy Majorana neutrinos will then
contribute to $\tilde A$.
For definiteness, we have restricted ourselves to
a model with $n_G=2$ (neglecting mixings due to  $\nu_{\tau}$),
where the mixings $B_{lN_\alpha}$ and the two heavy neutrino
masses, $m_{N_1}$ and $m_{N_2}$, satisfy the relation~\cite{KPS}
\beq
B_{lN_2}B_{l'N_2}^{*}=\frac{m_{N_1}}{m_{N_2}}B_{lN_1}B_{l'N_1}^{*}.
\label{B21}
\eeq 
The branching ratio for $K_L\to e \mu$ may conveniently be
calculated by using isospin invariance relations between
the decay amplitudes of $\bar{K}^0\to \mu^-e^+$ and
$K^-\to \mu^-\nu_\alpha$.
Setting $m_e=0$ relative to $m_\mu$ in the phase space,
one finds
\beq
B(K_L\to e\mu)=4.1\times10^{-4}\left| \tilde A\right| ^2.
\eeq  
Experimental bounds coming from the nonobservation of
the decay $\mu\to e\gamma$ will constrain the parameter
space of our theory and impose severe limits on the decay
$K_L\to e\mu$. In Majorana neutrino models, the branching
ratio of $\mu\to e\gamma$ is
\beq
B(\mu\to e\gamma )=\frac{6\alpha}{\pi}\Big|\sum_{\alpha=1,2}
 B_{\mu N_\alpha}B_{eN_\alpha}^{*}F(\la)\Big|^{2},\label{Bmueg}
\eeq 
where the loop function $F$ calculated in~\cite{Inami,Ma} is given by
\beq
F(\la)= \frac{2\la^{3}+5\la^{2}-\la}{4(1-\la)^3}
         +\frac{3\la^{3}\ln \la}{2(1-\la)^4}.
\eeq 
The present experimental upper limit $B(\mu\to e\gamma )<4.9\times
10^{-11}$ together with Eqs.~(\ref{B21}) and~(\ref{Bmueg})
can be used to obtain combined constraints on the mixings $B_{lN_\alpha}$
and heavy neutrino masses.
These constraints are quite useful in order to individually
evaluate the contribution of the charm and top quark to $B(K_L\to
e\mu)$. In our numerical analysis, we have used the maximally allowed values
$V_{td}=0.018$ and $V_{ts}=0.054$, and the central value
for the top-quark $m_t=175$~GeV as has recently been reported by the CDF
collaboration~\cite{CDF}.
Contrary to~\cite{Acker} where only one heavy neutrino family with mass not
much heavier than $M_W$ was considered, we find that the charm-quark
contribution is negligible and only top-quark quantum effects are of
interest here for heavy neutrino masses larger than 150 GeV.
Of course, the mass of the heavy neutrinos should not exceed an upper limit
that invalidates perturbative unitarity. This mass limit is qualitatively
estimated to be no bigger than 50 TeV~\cite{HN}. From Fig.~2,
we see that the branching ratio takes the maximum value
$B(K_L\to e\mu)=5.5\times 10^{-15}$, which is still rather far from
the present experimental sensitivity $B(K_L\to e \mu)<3.3\times 10^{-11}$
at $90\%$ C.L.~\cite{PDG}.
In Fig.~2, we have further assumed for the two heavy neutrinos, $N_1$ and
$N_2$, to have about the same mass $m_N$. Nevertheless, in Fig.~3, we
have plotted the dependence of the branching ratio as a function of the
value $\rho =m_{N_2}/m_{N_1}$  for selected values of $m_{N_1}$.
The solid curve in Fig.~3 determines an upper limit of the allowed
region for $B(K_L\to e \mu)$ by taking into account the combined
constraints arising from the neutrino mixings $B_{lN_\alpha}$ and
the validity of perturbative unitarity.

An enhancement of the decay rate can be obtained~\cite{PRD} by
considering a general left-right symmetric model based on the
gauge group $SU(2)_R\otimes SU(2)_L\otimes U(1)_{B-L}$.
This model predicts two charged gauge bosons $W_L$ and $W_R$,
which are generally not mass eigenstates, but the relevant mixing
angle is proportional to the vacuum expectation value ($v_L$) of
the left-handed Higgs triplet $\Delta_L$ with quantum numbers
$(0,1,2)$. For simplicity, we will work out the realistic case~(d)
of Ref.~\cite{Gunion}, in which $v_L=0$. In this case, the $W_L$ and
$W_R$ bosons become mass eigenstates with masses $M_L=M_W$ and $M_R$,
respectively.

In the context of left-right scenarios, there exist three different
sets of diagrams depending on the way that the virtual gauge bosons
$W_L$ and $W_R$ are involved. To be precise, we group the
four diagrams in Fig.~1(a)--(d) separately which are entirely mediated
by $W_L$ bosons and are identical with those considered above.
As a distinctive set, we consider the Feynman graphs in Fig.~1(e)--(l),
in which a $W_L$ and a $W_R$ boson are simultaneously present.
Furthermore, there are four graphs (not shown in Fig.~1) that can be
obtained from the first group by replacing the $W_L$ boson by the $W_R$ one.
In addition, at tree level, Higgs scalars with FCNC couplings should be
taken into account to cancel possible gauge-dependent terms
arising from the graphs in Fig.~1~\cite{IL}.
The inclusion of such Higgs-dependent graphs is not expected to
alter quantitatively the results obtained in our analysis.
For a discussion on related issues, the reader is referred
to Ref.~\cite{PRD,Gauge}.

Since the first set of graphs has already been considered in the context
of the SM with right-handed neutrinos, we will therefore proceed with the
calculation of the graphs of the second one, which are depicted in
Fig.~1(e)--(l). Their contribution to the corresponding reduced amplitude
is found to be
\bea
\tilde A_{LR} & = & \beta_g\eta\left\{
\begin{array}{c}
 \\
 \\
\end{array}\!\!\!\!
\sum_{i=c,t}V_{id}^{L*}V_{is}^{R}
\sum_{\alpha=n_G+1}^{2n_G}B_{\mu
\alpha}^{R}B_{e\alpha}^{L*}(\li\la)^{1/2}\right.
\nonumber\\
& &\times\left[ \left( 1+\frac{\beta\li\la}{4}\right) J_1(\li,\la,\beta)
- \frac{1+\beta}{4}J_2(\li,\la,\beta)\right] \nonumber\\
& & +
\sum_{i=c,t}V_{id}^{R*}V_{is}^{L}
\sum_{\alpha=n_G+1}^{2n_G}B_{\mu \alpha}^{L}B_{e\alpha}^{R*}(\li\la)^{1/2}
\nonumber\\ & &\left.
\left[ \left( 1+\frac{\beta\li\la}{4}\right) J_1(\li,\la,\beta)
- \frac{1+\beta}{4}J_2(\li,\la,\beta)\right]
\begin{array}{c}
  \\
  \\
\end{array} \!\!\!\!
\right\} .\label{ALR}
\eea 
Here, $\beta =M_{L}^{2}/M_{R}^{2}$, $\beta_g=(g_{R}^{2}/g_{L}^{2})
M_{L}^{2}/M_{R}^{2}$, with $g_L=g_w$ and $g_R$ being the coupling constants
related to the gauge groups $SU(2)_L$ and $SU(2)_R$, respectively. The
parameter $\eta$ in Eq.~(\ref{ALR}) is an enhancement factor that results
from the different type of operator describing the kaon-to-vacuum matrix
element. The factor $\eta$ is defined as~\cite{PRD}
\bea
\eta &=&
\frac{\left< 0\left| \overline{d}\gamma^{\sigma}\gamma^{\kappa}
\PR s\right| \overline{K}^{0}\right> \overline{u}_{\mu}\gamma_{\sigma}
\gamma_{\kappa}\PL v_{e}}{
\left< 0\left| \overline{d}\gamma^{\alpha}\PL s
\right| \overline{K}^{0}\right> \overline{u}_{\mu}
\gamma_{\alpha} \PL v_{e}}\nonumber\\
& \simeq & \frac{4M_{K}^{2}}{(m_s+m_d)m_{\mu}}\simeq 50\, ,
\eea 
which is estimated by using the assumption of partial conservation
of the axial vector current (PCAC ). The box functions $J_1$ and $J_2$ are
given by
\bea
J_1(\li,\la,\beta )&=&
\frac{\li\ln\li}{(1-\li)(1-\beta\li)(\la-\li)}+
\frac{\la\ln\la}{(1-\la)(1-\beta\la)(\li-\la)}\nonumber\\
&+&
\frac{\beta\ln\beta}{(1-\beta)(1-\beta\li)(1-\beta\la)}\  ,
\nonumber\\
J_2(\li,\la,\beta )&=&
\frac{\li^{2}\ln\li}{(1-\li)(1-\beta\li)(\la-\li)} +
\frac{\la^{2}\ln\la}{(1-\la)(1-\beta\la)(\li-\la)}\nonumber\\
&+&
\frac{\ln\beta}{(1-\beta)(1-\beta\li)(1-\beta\la)} \ .
\eea 
In Eq.~(\ref{ALR}), $B^L$ and $V^L$ are essentially
the matrices $B$ and $V$ which have been
defined above in the SM with right-handed neutrinos.
By analogy, the matrices $B^R$ and $V^R$ parametrize the interaction
of $W_R$ with leptons and quarks, respectively. In case of no
manifest or pseudomanifest left-right symmetry, there are no experimental
constraints on the elements of $V^R$ and they are limited simply by
unitarity. In fact, for specific forms of $V^R$ given in Table~II
of Ref.~\cite{PRD}, the $K_L-K_S$ mass difference imposes a lower
bound on the $W_R$-boson mass $M_R\apprge 400$~GeV, not very different
from the experimental one~\cite{PDG}.

Following a similar procedure, we can extract constraints on $B^R$
from the experimental limit of the decay $\mu\to e\gamma$ that can
be mediated by right-handed currents. In particular, if the mixings
$B^L_{\mu N_\alpha}$ are assumed to be extremely suppressed or vanish,
then only $W_R$ bosons can provide a nonzero value to the decay
$\mu\to e\gamma$. Therefore, in the evaluation of $B(K_L\to e\mu)$, we will
consider only the first term of the r.h.s.~of Eq.~(\ref{ALR}), which is
rather conservative. Then, the dominant contribution to the amplitude
originates from the diagram~(h) in Fig.~1. In the limit of
$m_N, M_R\gg M_L$, the reduced amplitude $\tilde{A}_{LR}$ behaves
asymptotically as
\beq
\tilde{A}_{LR}\ \approx\ (1.2\ 10^{-4})\times \eta\beta_g V_{td}
\left( \frac{s^{\nu_e}_L}{0.01}\right)
\frac{\beta^{1/2}\lt^{3/2}\lN^{1/2}}{4(1-\beta\lt)}
\left( \ln\beta\ +\ \frac{\lt\ln\lt}{\lt -1} \right),
\label{Alr}
\eeq 
where constraints for the mixing matrices $B^L$ and $B^R$
coming from the decay $\mu\to e\gamma$ have been
implemented.
In Eq.~(\ref{Alr}), we have identified $(s^{\nu_l}_L)^2\equiv
\sum\limits_{\alpha=1}^{n_G} |B^L_{lN_\alpha}|^2$, where $l$ stands
for an electron or a muon. In agreement with a global analysis of low-energy
and LEP data~\cite{CPB}, we have considered
$(s^{\nu_e}_L)^2=(s^{\nu_\mu}_L)^2 \leq 10^{-4}$ in our numerical
estimates. Nevertheless, one may have to worry that diagrams similar
to Fig.~1(h) and~1(l), which are present in the decay $\mu\to eee$,
could lead to a violation of the experimental bound
$B(\mu\to eee)<10^{-12}$~\cite{PDG}. Considering only the
dominant nondecoupling terms, we have estimated that this happens
when $\eta m_t V_{td}/(s^{\nu_e}_Lm_N)<1$ for $M_R\sim 1$ TeV.
Unless the mass of heavy neutrinos $m_N>10$~TeV for $M_R\apprle 1$~TeV,
the limits derived from the nonobservation of $\mu\to e\gamma$ will
be rather sufficient to preform our combined analysis.

As can be seen from Fig.~4, $B(K_L\to e\mu)$ depends strongly on
$M_R$ via the parameter $\beta$. As a natural choice, we have assumed
the left-right symmetric case $\beta=\beta_g$.
Taking the constraints coming from $\mu\to e\gamma$ into account,
we find that heavy neutrinos with few TeV masses can give rise to
branching ratios of the order of $10^{-11}$ close to the present
experimental limit. Note that there is a local
maximum in Fig.~4 for smaller values of $\beta$, where $W_R$ bosons
with several TeV masses can also account for
$B(K_L\to e\mu) \sim 10^{-11}$.

There is a third set of graphs (not shown
in Fig.~1) contained in the reduced amplitude $\tilde{A}_{RR}$,
in which a $W_L$ ($\chi_L$) boson should be replaced by a $W_R$
$(\chi_R)$ one in Fig.~1(a)--(d). From Eq.~(\ref{red}), it is
straightforward to obtain the analytic expression for $\tilde{A}_{RR}$
by making the obvious substitutions mentioned above. In this way, one has
\beq
\tilde{A}_{RR}=\beta_g^2 \sum_{i=c,t} V_{id}^{R*}V_{is}^{R}
\sum_{\alpha=n_G+1}^{2n_G}B_{\mu j}^{R}B_{ej}^{R*} E(\beta\li,\beta\la).
\label{ARR}
\eeq
{}From Eq.~(\ref{ARR}), we find numerically that $|\tilde{A}_{RR}|^2$
is about one order of magnitude smaller than $|\tilde{A}_{LR}|^2$ due
to the severe constraints coming from the nonobservation of
$\mu\to e\gamma$. For instance, for $M_R=800$ GeV and $m_N=20$ TeV,
the branching ratio
$B(K_L\to e\mu)$ is here $1.6\times 10^{-12}$ as compared to the
value $B(K_L\to e\mu)=2.0\times 10^{-11}$ in a complete computation.

In conclusion, we have analyzed some interesting aspects of the
decay $K_L\to e\mu$ in the framework of Majorana-neutrino models
based on the gauge groups $SU(2)_L\otimes U(1)_Y$ and
$SU(2)_R\otimes SU(2)_L\otimes U(1)_{B-L}$. In comparison to a
previous work~\cite{PRD}, we wish to stress that in the renormalizable gauge,
diagrams with would-be Goldstone bosons are indeed important, since
the stringent limits on the mixings between light and heavy neutrinos
can be relaxed by the presence of two heavy neutrino families.
In an $SU(2)_L\otimes U(1)_Y$ model with right handed neutrinos,
we have found that $B(K_L\to e\mu)\apprle 10^{-15}$
for heavy neutrinos with TeV masses, where the top-quark contribution
prevails over the charm-quark one.
In an $SU(2)_R\otimes SU(2)_L\otimes U(1)_{B-L}$ model with Majorana
neutrinos, the chirality changing diagrams (h) and (l) in Fig.~1
dominate in the branching ratio over the remaining set of graphs.
Through heavy-neutrino-chirality enhancements
the resulting branching ratio $B(K_L\to e\mu)$ can be as large as
the present experimental limit $\sim 10^{-11}$, for a wide
range of parameter values. In addition, the constraints derived from
$B(K_L\to e\mu)$ are complementary to the ones determined by other
low-energy experiments, such as the possible decays $\mu\to e\gamma$
and $\mu\to eee$, and the $K_L-K_S$ mass difference. Therefore,
experimental tests at DA$\Phi$NE or in other kaon factories will be
very crucial and may reveal surprises in the leptonic decay channels
of the $K_L$ meson that might signal the onset of new physics.\\[0.7cm]
{\bf Acknowledgements.} Helpful discussions with Jose Bernab\'eu,
Francisco Botella, and Amon Ilakovac are gratefully acknowledged.

\newpage

\newpage

\centerline{\bf\Large Figure Captions }
\vspace{-0.2cm}
\newcounter{fig}
\begin{list}{\bf\rm Fig. \arabic{fig}: }{\usecounter{fig}
\labelwidth1.6cm \leftmargin2.5cm \labelsep0.4cm \itemsep0ex plus0.2ex }

\item Feynman diagrams contributing to $K_L \to e \mu$ in Majorana
      neutrino models relying on the gauge groups:
      (a)--(d) $SU(2)_L\otimes U(1)_Y$ and (a)--(l)
      $SU(2)_R\otimes SU(2)_L\otimes U(1)_{B-L}$.

\item $B(K_L\to e\mu)$ as a function of the heavy neutrino
      mass $m_N\ (\simeq m_{N_1} \simeq m_{N_2})$
      ($m_t=175$~GeV) in the SM with right-handed neutrinos.

\item $B(K_L\to e\mu)$ versus $\rho =m_{N_2}/m_{N_1}$
      in the $SU(2)_L\otimes U(1)_Y$ model with neutral singlets.

\item $B(K_L\to e\mu)$ as a function of $\beta=M^2_L/M^2_R=\beta_g$
      in an $SU(2)_R\otimes SU(2)_L\otimes U(1)_{B-L}$ model, assuming
      that all heavy neutrinos are approximately degenerate
      with mass $m_N$.

\end{list}

\end{document}